\documentclass[12pt]{article}
\usepackage{jheppub}
\usepackage{mathtools,amssymb,amsthm}
\usepackage[utf8]{inputenc}
\usepackage{enumitem}
\usepackage{xcolor}
\usepackage{graphicx}
\usepackage{multirow}
\usepackage[normalem]{ulem}
\usepackage{longtable}
\usepackage{tikz}
\allowdisplaybreaks

\newcommand{\be}{\begin{equation}}
\newcommand{\ee}{\end{equation}}
\newcommand{\bi}{\begin{itemize}}
\def \bea {\begin{eqnarray}}
\def \eea {\end{eqnarray}}

\def\ba#1\ea{\begin{align}#1\end{align}}
\def\bad#1\ead{\begin{aligned}#1\end{aligned}}
\def\bg#1\eg{\begin{gather}#1\end{gather}}
\def\bm#1\em{\begin{multline}#1\end{multline}}
\def\bmd#1\emd{\begin{multlined}#1\end{multlined}}

\newcommand{\ignore}[1]{}

\def \bal#1\eal  {\begin{align} #1 \end{align}}
\def \bga#1\ega  {\begin{gather} #1 \end{gather}}
\def\({\left(}
\def\){\right)}
\def\[{\left[}
\def\]{\right]}
\def\<{\left\langle}
\def\>{\right\rangle}

\newcommand{\eim}{\end{itemize}}
\newcommand{\beq} {\begin{equation}}
\newcommand{\eeq} {\end{equation}}
\newcommand{\bc}{\begin{center}}
\newcommand{\ec}{\end{center}}

\begin{document}

\title{Wheeler DeWitt States of a Charged AdS$_4$ Black Hole}

\author[a]{Matthew J. Blacker}\author[b]{ and Sirui Ning}
\affiliation[a]{Department of Applied Mathematics and Theoretical Physics,
University of Cambridge, Cambridge CB3 0WA, UK}
\emailAdd{mjb318@cam.ac.uk, sirui.ning@physics.ox.ac.uk}
\affiliation[b]{The Rudolf Peierls Centre for Theoretical Physics, University of Oxford,\\Oxford OX1 3PU, UK}

\abstract{We solve the Wheeler DeWitt equation for the planar Reissner-Nordström-AdS black hole in a minisuperspace approximation. We construct semiclassical Wheeler DeWitt states from Gaussian wavepackets that are peaked on classical black hole interior solutions. By using the metric component $g_{xx}$ as a clock, these states are evolved through both the exterior and interior horizons. Close to the singularity, we show that quantum fluctuations in the wavepacket become important, and therefore the classicality of the minisuperspace approximation breaks down. Towards the AdS boundary, the Wheeler DeWitt states are used to recover the Lorentzian partition function of the dual theory living on this boundary. This partition function is specified by an energy and a charge. Finally, we show that the Wheeler DeWitt states know about the black hole thermodynamics, recovering the grand canonical thermodynamic potential after an appropriate averaging at the black hole horizon.}

\maketitle


\section{Introduction}
In recent decades, the holographic principle has made it possible to study the bulk dynamics of gravitational theories via processes in a dual quantum theory. This dual quantum theory lives on some slice, usually the boundary, of the spacetime. For asymptotically AdS (anti de Sitter) spacetimes, an important part of the holographic toolbox has been the holographic renormalization group flow. This tells us that events further from the boundary correspond to lower energy processes in the dual theory \cite{susskind_holographic_1998,skenderis_lecture_2002}. The location in the bulk can be quantified by the metric function $g_{tt}$. Specifically, the AdS boundary lives at $g_{tt} \rightarrow -\infty$ and corresponds to the ultraviolet limit of the dual theory, and moving to lower energy processes corresponds to increasing $g_{tt}$. Reaching a horizon in the spacetime corresponds to $g_{tt} \rightarrow 0$, or equivalently the far infrared of the renormalization group. At this point, the usual interpretation from the dual theory perspective is that there are no more modes to integrate out.\\\\
However, it was recently emphasised in \cite{Frenkel:2020ysx} that it is natural to extend the renormalization group flow in the bulk through the horizon, including as far as the singularity. Subsequent work has further explored extending the renormalizaton group flow through the horizon in various setups \cite{Wang:2020nkd,Caceres:2021fuw,Mansoori:2021wxf,Bhattacharya:2021nqj,Das:2021vjf,Caceres:2022smh}, including when the black hole has charge \cite{Hartnoll:2020fhc,Hartnoll:2020rwq,Sword:2021pfm}. The upshot of these developments is that the renormalization group flow might be added to our current list of tools \cite{fidkowski_black_2004-1,hartman_time_2013-1,stanford_complexity_2014-1,brown_holographic_2016} for probing black hole interiors. What remains to be achieved is using the flow to explicitly construct the interior from the exterior.\\\\
A hint on how to do this is offered by the observation that the natural language for describing the renormalization group flow is via the Hamilton-Jacobi equation of classical mechanics \cite{deBoer:1999tgo,Faulkner:2010jy,Heemskerk:2010hk}. The Hamilton-Jacobi equation arises in the context of quantum cosmology as the classical limit of the Wheeler DeWitt equation \cite{Halliwell:1987eu,Halliwell:1989myn}. Furthermore, the Wheeler DeWitt equation arises from the canonical quantization of the Einstein-Hilbert action so is a natural tool for studying quantum aspects of spacetime. This point has been emphasised recently in \cite{chowdhury_holography_2022,witten_note_2022}. Therefore, Wheeler DeWitt states constructed from Hamilton-Jacobi functions are a probe of semiclassical aspects of spacetime that naturally connect with the holographic renormalization group flow. They may therefore offer a window into the black hole interior. \\\\
This was indeed the approach of \cite{Hartnoll:2022snh}, which studied how the (semiclassical) quantum state of the interior of an AdS-Schwarzschild black hole is prepared by the AdS boundary. In that work, the interior wavefunction was identified with the boundary partition function extended to $g_{tt} > 0$. This was achieved by constructing Wheeler DeWitt states from solutions to the Hamilton-Jacobi equation for a number of different clocks. The most convenient of these solutions was linear in $g_{tt}$, so it was natural to extend from $g_{tt} < 0$ to $g_{tt} >0$ when passing through the horizon. Indeed, in the setup of \cite{Hartnoll:2022snh} $g_{tt}$ was monotonic and a well-defined clock from the boundary to the singularity. When a non-monotonic clock such as volume is used, solutions to the Wheeler-DeWitt equation involve, for example, superpositions of spacetimes close to the singularity and close to the horizon -- both of which have small spatial volumes -- and are hence harder to interpret. Non-monotonic clocks are therefore a less convenient choice for probing the full space time.\\\\
What happens, then, if $g_{tt}$ is not monotonic? This arises in spacetimes with more than one horizon. It is natural to ask what choice of clock would be best for applying the framework of \cite{Hartnoll:2022snh} to these spacetimes. One such case is the Reissner–Nordström-AdS (RN AdS) spacetime, where the charge of the black hole introduces a second (Cauchy) horizon in the black hole interior. \begin{figure}[h!]
\centering
\includegraphics[width=.45\textwidth]{Figure1}
\caption{The regions of the Penrose diagram of the RN-AdS spacetime that will play a role in our discussion. The black dotted lines indicate the black hole and cauchy horizons, where $g_{xx}= g_{xx,+}$ and $g_{xx} = g_{xx,-}$ respectively. At the AdS Boundary $g_{xx} = \infty$ and at the singularity $g_{xx} = 0$. The value of $g_{tt}$ at each of these points is indicated to emphasise that it is not monotonic. We solve the Wheeler DeWitt equation in the black hole interior on constant $g_{xx}$ slices, and extend these to the cauchy interior in Section \ref{SectionClocks} and the exterior in Section \ref{SecPartitionFunction}.\label{FigPenrose}}
\end{figure} \\\\
Exploring the procedure of \cite{Hartnoll:2022snh} in the (planar) RN AdS spacetime is the aim of this work. As is emphasised in Figure \ref{FigPenrose}, in the RN AdS spacetime $g_{tt}$ is not monotonic from the AdS boundary to the singularity, whereas $g_{xx}$ is. We therefore solve the Wheeler DeWitt equation in the black hole interior, and use $g_{xx}$ as a clock to extend these solutions to the cauchy interior and the exterior. In the cauchy interior, we study whether these states can probe the singularity. In the exterior, we identify the Wheeler DeWitt states with the partition function of the dual theory living on the AdS boundary via a renormalization group flow in $g_{xx}$.\\\\
This paper is organised as follows. In Section \ref{SecRNAdSHJ}, we recover the RN-AdS black hole from Hamilton-Jacobi equation. Using the corresponding Hamilton-Jacobi function, we construct semiclassical Wheeler DeWitt states in Section \ref{SecSemiClassWdW}. We then show three main results. Firstly, in Section \ref{SectionClocks} we use $g_{xx}$ as a clock to show that the variance in $\langle g_{tt} \rangle$ blows up as the singularity is approached. This result indicates that minisuperspace classicality breaks down. Secondly, in Section \ref{SecPartitionFunction} we show that the Wheeler DeWitt states are the continuation of the boundary partition function along $g_{xx}$ slices. That is, the Wheeler DeWitt states are part of the renormalization group flow of a dual theory living on the AdS boundary. The dual theory is specified by an energy and a charge. Finally, in Section \ref{SecThermodynamics} we show that the Wheeler DeWitt states know about the thermodynamics of the black hole horizon. We do so by averaging over states of fixed extrinsic curvature at the black hole horizon in the spirit of \cite{Blacker:2023oan}, to recover the grand canonical thermodynamic potential.
\section{Reissner-Nordström-AdS from the Hamilton-Jacobi equation}
\label{SecRNAdSHJ}
As will be elaborated upon in Section \ref{SecSemiClassWdW}, the classical limit of the Wheeler-DeWitt equation is the Hamilton-Jacobi equation for the Einstein-Hilbert action. These classical solutions can then be used to construct semi-classical wavepackets. We will thus begin by recovering the RN-AdS black hole using the Hamilton-Jacobi formulation of gravity, following the approach of \cite{Blacker:2023oan,Hartnoll:2022snh}. \\\\
The action for four dimensional gravity with a negative cosmological constant and a Maxwell field is
\begin{align}
    I\left[ g, A \right]= \int d^4x\sqrt{-g}(R+6)-\frac{1}{4}\int d^4x \sqrt{-g} F^2 + 2 \int d^3x \sqrt{h}K .
\label{EqAction}
\end{align}
Here, we work in units $16 \pi G_N = 1$. The second term in \eqref{EqAction} is the Gibbons-Hawking boundary term, and the final term is the action of a Maxwell field of strength $F = dA$ for a vector potential $A$. Throughout this paper, we will consider the ans\"{a}tze
    \begin{align}
    ds^2 = - N^2 dr^2 + \frac{g_{tt}}{\left( \Delta t \right)^2} dt^2 + \frac{g_{xx}}{\left( \Delta x \right)^2} \left( dx^2 + dy^2 \right), \text{ and } A = \frac{\phi_t}{\Delta t} dt,
\label{EqAnsatz}
\end{align}
where $N$, $g_{tt}$, $g_{xx}$ and $\phi_t$ are functions of $r$ only. As in \cite{Blacker:2023oan}, we explicitly include in the metric factors of $\Delta t$ and $\Delta x$, the extent of the $t$ and $\lbrace x, y \rbrace$ coordinates respectively. These factors simply re-scale our metric functions in a way that is convenient for the rest of this work.\\\\
We are interested classically in the black hole interior, as depicted in Figure \ref{FigPenrose}. On the ansatz \eqref{EqAnsatz}, the action is
\begin{align}
    I [N, g_{tt}, g_{xx}, \phi_t] = \int dr \mathcal{L},
\end{align}
where the Lagrangian density is
\begin{align}
    \mathcal{L} = \frac{g_{xx}^2 \left(\left(\partial_r \phi_t \right)^2+(12/L^2) g_{tt} N^2\right)-2 g_{xx} \partial_r g_{tt} \partial_r g_{xx}-g_{tt} \left( \partial_r g_{xx} \right)^2}{2 \sqrt{g_{tt}} g_{xx} N}.
\label{EqEulerLagrange}
\end{align}
From this action we define the momenta $\pi_i = \lbrace \pi_{tt}, \pi_{xx}, \pi_{\phi} \rbrace$ as conjugate to $g_i = \lbrace g_{tt}, g_{xx}, \phi_t \rbrace$, and construct a Hamiltonian in the usual way. $N$ plays the role of a Lagrange multiplier, and imposes the Hamiltonian constraint 
\begin{align}
    -\frac{1}{2} g_{tt} \pi_{tt}^2 +g_{xx} \pi_{tt} \pi_{xx}+ \frac{6}{L^2} g_{xx}^2-\frac{\pi_{\phi}^2}{2} = 0.
\label{EqHamiltonConstraint}
\end{align}
To reconstruct the Hamilton-Jacobi equation, we introduce the Hamilton-Jacobi function $S(g_i)$ such that $\pi_{i} = \partial_{g_i} S(g_i)$. From \eqref{EqHamiltonConstraint} we have
\begin{align}
    g_{xx} \left( \partial_{g_{xx}} S \right) \left( \partial_{g_{tt}} S \right) -\frac{1}{2} \left( \partial_{\phi_t} S \right)^2-\frac{1}{2} g_{tt} \left( \partial_{g_{tt}} S \right)^2+ \frac{6}{L^2} g_{xx}^2 = 0.
\label{EqHJEquation}
\end{align}
The classical equations of motion are obtained from the Hamilton-Jacobi equation by first finding one member of the family of solutions to \eqref{EqHJEquation}. As there are three variables $g_i$, we expect the solution to have three constants of integration. Two of these are non-trivial, and the third leads to an overall shift in in $S$ which does not contribute to the equations of motion and we therefore do not consider. The first such solution with which we will be concerned is
\begin{align}
    S_1 \left( g_{tt}, g_{xx}, \phi_t; k_0, c_0 \right) = -\frac{e^{-k_0} \left(c_0^2+4 g_{xx}^2/L^2\right)}{2 \sqrt{g_{xx}}}+c_0 \phi_t+2 g_{tt} \sqrt{g_{xx}} e^{k_0},
\label{solution}
\end{align}
where $\lbrace k_0, c_0 \rbrace$ are the non-trivial constants of integration. The classical solution (i.e. the solution to Euler-Lagrange equations corresponding to \eqref{EqEulerLagrange}) is obtained by introducing another pair of constants $\lbrace \epsilon, \mu \rbrace$ such that
\begin{align}
    \partial_{k_0} S_1 = \epsilon, \text{ and } \partial_{c_0} S_1 = \mu.
\label{EqClassicalconstraints}
\end{align}
Solving this pair of equations we can rewrite two of our coordinates $\lbrace g_{tt}, \phi_t \rbrace$ in terms of the other and the constants of integration, as
\begin{align}
    g_{tt} = \frac{e^{-2 k_0} \left(-c_0^2 L^2 -4 g_{xx}^2+2 \sqrt{g_{xx}} L^2 e^{k_0} \epsilon \right)}{4 L^2 g_{xx}}, \text{ and } \phi_t - \mu = \frac{c_0 e^{-k_0}}{\sqrt{g_{xx}}}.
\label{EqClassicalsol}
\end{align}
The upshot of \eqref{EqClassicalsol} is that we can now recover the RN-AdS solution. Substituting \eqref{EqClassicalsol} into the equation of motion for $N$ from \eqref{EqEulerLagrange}, we obtain
\begin{align}
    N^2 dr^2 = - \frac{dg_{xx}^2}{c_0^2 - 2e^{k_0}\epsilon \sqrt{g_{xx}} + 4 g_{xx}^2/L^2}.
\label{EqNsquaredonshell}
\end{align}
Finally, it will be convenient to define $z = 1/\sqrt{g_{xx}}$, so that when substituting \eqref{EqNsquaredonshell} into \eqref{EqAnsatz} we obtain
\begin{align}
    ds^{2}=\frac{1}{z^2} \left(-f(z)e^{-2k_0} 
\frac{dt^2}{(\Delta t )^2} +\frac{dz^2}{f(z)}+\frac{1}{(\Delta x)^2} \left( dx^2+dy^2 
 \right)  \right),\text{ and } A=\frac{\Phi(z)}{\Delta t}dt,
\label{EqRNAdS}
\end{align}
where
\begin{align}
    f(z)=\left( \frac{1}{L^2} -\frac{z^3}{2}\epsilon e^{k_0}+\frac{c_0^2}{4}z^4\right),\text{ and } \Phi(z)=\mu+c_0 e^{-k_0}z.
\label{EqFZ}
\end{align}
We recognise \eqref{EqRNAdS} as the RN-AdS solution. The solution has a Cauchy and black hole horizon where $f(z_h) = 0$. The function $f(z) > 0$ in the black hole exterior where $t$ is the timelike direction, changes sign to $f(z) < 0$ after crossing the black hole horizon, and changes sign again to $f(z) > 0$ in the interior of the Cauchy horizon. In the coordinate $z$, we have that (on-shell)
\begin{align}
    g_{tt} = - \frac{f(z)}{z^2} e^{-2k_0}, \text{ and } \phi_t - \mu = c_0e^{-k_0} z.
\label{EqClassicalsolinz}
\end{align}
Therefore, a consequence of having two horizons is that $g_{tt}$ is only monotonic when restricted to either $z > z^*$ or $z < z^*$, where $z^* = 3\epsilon e^{k_0}/2c_0^2$. This means that $g_{tt}$ is not a suitable candidate for a clock to define a relational notion of time. However, $g_{xx}$ is trivially monotonic, and from \eqref{EqClassicalsolinz} we can see that $\phi_t$ is also monotonic, and so are both alternative choices of clock. This point will be discussed further in Section \ref{SectionClocks}. \\\\
The radial functions should satisfy the well-known asymptotic properties of charged black hole solutions to Einstein-Maxwell-AdS theory \cite{hartnoll_holographic_2018}. To recognise our solution with these, it is instructive for a moment to consider the re-scaled time coordinate $t' = e^{-k_0} t/\Delta t$ where
\begin{align}
    ds^{2}=\frac{1}{z^2} \left(-f(z)dt'^2+\frac{dz^2}{f(z)}+\frac{1}{(\Delta x)^2} \left( dx^2+dy^2 
 \right)\right),\text{ and } A=\Phi(z)e^{k_0} dt'. 
\label{EqRNAdS}
\end{align}
We therefore expect to recover from the electromagnetic potential the boundary chemical potential $\mu_B$ 
\begin{align}
    \lim_{z\rightarrow 0} \Phi(z) e^{k_0} = \mu_B,
\end{align}
and the boundary charge density $\rho$
\begin{align}
    \rho = - \lim_{z\rightarrow 0} \partial_z \Phi(z) e^{k_0}.
\end{align}
In particular, for the classical solution \eqref{EqClassicalsolinz} we recover
\begin{align}
    \mu_B = \mu e^{k_0} \text{ and } \rho = -c_0.
\label{EqMuandRho}
\end{align}
We can compute the associated charge $Q$ via the four-current $J_{\mu}$ and associated four vector $n^{\mu}$ on an $r$ slice
\begin{align}
    Q = \int d^3 x \sqrt{-g} J_a n^a = \rho e^{-k} = -c_0 e^{-k_0}.
\label{EqChargeDefinition}
\end{align}
We can additionally fix the value of $\mu$ (or equivalently $\mu_B$). In the RN-AdS geometry, if one rotates to Euclidean signature only the outer horizon remains and the thermal circle shrinks to zero there \cite{hartnoll_holographic_2018}. Thus, for a Wilson loop of the Maxwell field to be regular, we require $\Phi(z_+) = 0$, where $z_+$ is the outer horizon. This regularity condition in the bulk fixes the chemical potential of the boundary theory to
\begin{align}
    \mu_B = - c_0 z_{+},
\label{EqFixMuB}
\end{align}
where $z_+$ is the value of $z = 1/\sqrt{g_{xx}}$ at the black hole horizon.\\\\
So far, we have only considered one member of the family of solutions to \eqref{EqHJEquation}. We could have, of course, constructed a classically equivalent solution where $\lbrace \epsilon, \mu \rbrace$ at the constants of integration. In fact, provided that we do not use both elements of a conjugate pair $\lbrace k_0, \epsilon \rbrace$ or $\lbrace c_0, \mu \rbrace$, we could construct a solution using any combination of these constants. In particular, we could construct the three following solutions
\begin{subequations}
    \begin{align}
        S_{\epsilon,c_0} = & c_0 \phi_t + \frac{F_1 - \epsilon^2 + \epsilon \sqrt{\epsilon^2 - F_1}}{\sqrt{\epsilon^2 - F_1} - \epsilon} - \epsilon \log \frac{\epsilon - \sqrt{\epsilon^2 - F_1}}{4g_{tt}\sqrt{g_{xx}}}, \label{EqSEpsc} \\
        S_{\epsilon,\mu} = & - F_2 - \epsilon \log \frac{4g_{xx}^{3/2}}{L^2(F_2 + \epsilon)}, \label{EqSEpsF} \\
        S_{k_0,\mu} = & \frac{e^{-k_0} \sqrt{g_{xx}}}{2} \left( e^{2k_0} \left( \left( \mu - \phi_t \right)^2 + 4g_{tt} \right) - \frac{4 g_{xx}}{L^2} \right). \label{EqSkf}
    \end{align}
\label{EqOtherHJsolutions}
\end{subequations}
Here, it has been convenient to define
\begin{subequations}
    \begin{align}
        F_1 \left(g_{tt},g_{xx};c\right) = & 4 g_{tt} \left( c^2 + 4 g_{xx}^2/L^2 \right), \\
        F_2 \left(g_{tt},g_{xx},\phi_t; \epsilon, f \right) = & \sqrt{\epsilon^2 - 4 \left( \left( \mu - \phi_t \right)^2 + 4g_{tt} \right) g_{xx}^2/L^2}.
    \end{align}
\end{subequations}
We have labelled each solution by the constants of integration it depends on. We wish to emphasise again here that each of \eqref{EqSEpsc}, \eqref{EqSEpsF}, and \eqref{EqSkf} is classically equivalent to $S_1$, in the sense that they lead to the same general solutions to the equations of motion. We have chosen to first focus on $S_1$ because in Section \ref{SecSemiClassWdW} it can be used to construct exact solutions to the Wheeler DeWitt equation. However, the relevance of each of \eqref{EqOtherHJsolutions} will become apparent as we proceed.
\section{Semiclassical Wheeler-DeWitt States}
\label{SecSemiClassWdW}
\subsection{From classical to semiclassical}
\label{SecClasstoSemiClass}
We can use these classical results to construct semiclassical quantum states, following a procedure we briefly review below. For a more detailed introduction to the procedure which follows, the interested reader should consult \cite{Halliwell:1987eu,Halliwell:1989myn}.\\\\
For the gravitational degrees of freedom, we have an action of the form
\begin{align}
    I[g] = \int dr \mathcal{L} = \int dr N \left[ \frac{1}{2N^2} G_{ab} \dot{g}^a \dot{g}^b - V(q) \right],
\end{align}
where $G_{ab}(g)$ are the minisuperspace metric and $V(g)$ an effective potential respectively, and are functions of some metric functions $g$. In the quantum theory, the Hamiltonian constraint is promoted to the Wheeler-DeWitt equation for a wavefunction $\Psi(g)$
\begin{align}
    \left( - \frac{1}{2} \nabla^2 + V(g) \right) \Psi \left( g \right) = 0,
\end{align}
where $\nabla^2$ is the Laplacian on the inverse DeWitt metric $ \sqrt{-G} \nabla^2 = \partial_{g^a} \left( \sqrt{-G} G^{ab} \partial_{g^b} \right)$. Strictly, there is an ordering ambiguity in quantizing the Hamiltonian constraint, which we discuss momentarily. If we restore units and consider solutions of the form $\Psi = \exp \left( i S(q)/\hbar \right)$, we find 
\begin{subequations}
    \begin{align}
        \mathcal{O} \left( \hbar^0 \right): & \left( \frac{1}{2} \left( \nabla S \right)^2 + V \right) \Psi = 0, \\
        \mathcal{O} \left( \hbar^1 \right): & \nabla^2 S \Psi = 0.
    \end{align}
\end{subequations}
Observe that the $\mathcal{O} \left( \hbar^0 \right)$ constraint is nothing more than the Hamilton-Jacobi equation. Moreover, the ordering ambiguity in defining the Laplacian only arises at order $\mathcal{O} \left( \hbar^1 \right)$. Thus, if we are only concerned with the leading order semiclassical physics, we can form a basis of states by exponentiating the Hamilton-Jacobi function.\\\\
In addition to the gravitational degrees of freedom, we also have electromagnetic terms, which amount to an additional term proportional to $\left( \partial_r \phi_t \right)^2$ in our action. In this case the above reasoning still holds, with the $\mathcal{O} \left( \hbar^0 \right)$ equation of motion still being the Hamilton-Jacobi equation and the $\mathcal{O} \left( \hbar^1 \right)$ equation containing an additional $\partial^2_{\phi_t} S$ term.
\subsection{Constructing states of the charged black hole}
\label{SecConstructingStates}
With the above discussion in mind, in our setup the Wheeler-DeWitt equation becomes
\begin{align}
    \frac{\partial}{\partial g_{tt}} \left( g_{tt} \frac{\partial \Psi}{\partial g_{tt}} - 2 g_{xx} \frac{\partial \Psi}{\partial g_{xx}} \right) + \frac{\partial^2 \Psi}{\partial \phi_t^2} + \frac{12}{L^2} g_{xx}^2 \Psi = 0.
\label{EqWdWEquation}
\end{align}
From section \ref{SecClasstoSemiClass}, we know that $e^{\pm i S}$ form a basis of semiclassical solutions to \eqref{EqWdWEquation} if $S$ is a solution to \eqref{EqHJEquation}. In fact, for $S_1$ \eqref{solution}, the solution
\begin{align}
    \Psi \left( g_{tt},g_{xx},\phi_t; k,c \right) = e^{iS_1\left( g_{tt},g_{xx},\phi_t; k,c \right)},
\label{EqBasisS1}
\end{align}
is an exact solution to the Wheeler DeWitt equation \eqref{EqWdWEquation}. Note that for the other Hamilton-Jacobi functions in \eqref{EqOtherHJsolutions}, $e^{\pm i S}$ are only solutions to leading order. We will thus first consider the basis of exact solutions to \eqref{EqWdWEquation} constructed from $S_1$. From the basis \eqref{EqBasisS1}, we construct a general solution
\begin{align}
    \Psi \left( g_{tt}, g_{xx}, \phi_t \right) = \int_{-\infty}^{\infty} \frac{dk}{2\pi} \int_{-\infty}^{\infty} \frac{dc}{2\pi} \beta \left( k,c \right) e^{iS_1\left( g_{tt},g_{xx},\phi_t; k,c \right)}.
\label{EqS1Solution}
\end{align}
Here $\beta \left( k, c \right)$ is an arbitrary function. By letting our constants of integration $\lbrace k, c \rbrace$ have either positive or negative sign, we only need to consider the $+iS_1$ solution.\\\\
As in \cite{Blacker:2023oan,Hartnoll:2022snh}, we can obtain another set of semiclassical solutions by fourier transforming to the basis corresponding to the classical constants $\lbrace \epsilon, \mu \rbrace$. We define
\begin{align}
    \beta \left( k, c \right) = \int \int \frac{d\epsilon}{\sqrt{2\pi}} \frac{d\mu}{\sqrt{2\pi}} \alpha \left( \epsilon, \mu \right) e^{- i \epsilon k } e^{-i \mu c},
\end{align}
where $\alpha \left( \epsilon, \mu \right)$ is an arbitrary function. The solution \eqref{EqS1Solution} then becomes
\begin{align}
    \begin{split}
        \Psi \left( g_{tt}, g_{xx}, \phi_t \right) = \int \frac{d\epsilon}{2\pi} \int\frac{d\mu}{2\pi} \alpha \left( \epsilon, \mu \right) & \frac{2\left( - g_{xx}/L^2 \right)^{1/2 + i\epsilon/2}}{\sqrt{2\pi}} \left( \frac{\left( \mu + \phi_t \right)^2 + 4 g_{tt}}{4} \right)^{-\frac{1}{2} \left( \frac{1}{2} + i \epsilon \right)} \\
        & \times K_{\frac{1}{2}+i \epsilon} \left( \frac{2 \sqrt{(\mu+\phi_t)^2 + 4g_{tt}} g_{xx}}{L} \right),
    \end{split}
\label{EqFourierSolution}
\end{align}
where $K_{1/2+i\epsilon}$ is a modified bessel function of the second kind. To make use of this Fourier transform in the semiclassical limit, we want to relate it to a solution of the Hamilton-Jacobi equation. To do so, we evaluate the Fourier transform using a stationary phase approximation and obtain the sum of two solutions
\begin{align}
    \Psi \left( g_{tt}, g_{xx}, \phi_t \right) = \int \frac{d\epsilon}{2\pi} \int\frac{d\mu}{2\pi} \alpha \left( \epsilon, f \right) \left[ \psi_{+} \left( g_{tt}, g_{xx}, \phi_t; \epsilon, \mu \right) + \psi_{-} \left( g_{tt}, g_{xx}, \phi_t; \epsilon, \mu \right) \right].
\label{EqPsiinEF}
\end{align}
These solutions are
\begin{align}
    \psi_{\pm} \left( g_{tt}, g_{xx}, \phi_t; \epsilon, \mu \right) = \frac{2g_{xx}}{L \sqrt{ F_2^2 \pm \epsilon F_2 }} \exp \left \lbrace i \left( \pm S_{\pm \epsilon,\mu} + \frac{\pi}{2} \right) \right \rbrace,
\end{align}
where $S_{\pm \epsilon,f}$ is the solution to the Hamilton-Jacobi equation introduced in \eqref{EqSEpsF}. As expected, taking the limit of no charge (i.e. the AdS-Schwarzschild solution), one finds that $\psi_{\pm}$ recover the solutions obtained in \cite{Hartnoll:2022snh}. As in \cite{Hartnoll:2022snh,Blacker:2023oan}, we only consider $e^{iS_{+\epsilon,\mu}}$ to ensure we have positive norm solutions in the interior. That is, in the semiclassical regime, we will find it useful to consider states
\begin{align}
     \Psi_{+} \left( g_{tt}, g_{xx}, \phi_t \right) = \int \frac{d\epsilon}{2\pi} \int\frac{d\mu}{2\pi} \alpha \left( \epsilon, \mu \right) e^{iS_{+ \epsilon, \mu}}.
\end{align}
\subsection{Gaussian wavepackets}
We are yet to consider a particular form for $\beta \left( k, c \right)$. As in \cite{Blacker:2023oan,Hartnoll:2022snh}, it is natural to consider Gaussian wavepackets, because these are strongly supported on the classical solution. In this work, we consider a Gaussian wavepacket
\begin{align}
    \begin{split}
        \beta \left( k, c \right) = & N_{\beta} \exp \left \lbrace -i \epsilon_0 \left( k - k_0 \right) - \frac{\Delta_k^2}{2} \left( k - k_0 \right)^2 \right \rbrace \\
        & \times \exp \left \lbrace -i \mu_0 \left( c - c_0 \right) - \frac{\Delta_c^2}{2} \left( c - c_0 \right)^2 \right \rbrace,
    \end{split}
\label{EqGaussianWavepacket}
\end{align}
where $N_{\beta} = \Delta_c \Delta_k/4\pi$. These wavepackets are strongly peaked around $\lbrace k = k_0, c = c_0 \rbrace$ if $\Delta_c, \Delta_k \gg 1$. In that case, we can then perform the $k$ and $c$ integrals in \eqref{EqS1Solution} by a 2D stationary phase approximation. The wavefunction is strongly peaked on values of the metric function where
\begin{align}
    \left. \frac{\partial S_1}{\partial k} \right|_{k=k_0} = \epsilon_0, \text{ and } \left. \frac{\partial S_1}{\partial c}\right|_{c=c_0} = \mu_0.
\end{align}
That is, the wavefunction is strongly supported on the classical solution \eqref{EqClassicalconstraints} with $\epsilon = \epsilon_0$ and $\mu = \mu_0$. As when computing the fourier transform, we have a contribution from two stationary points (corresponding to $S_{\pm \epsilon, \mu}$). As noted above, we only consider the branch corresponding to $S_{+\epsilon,\mu}$ to ensure positivity of the norm. Therefore, to leading semiclassical order the wavefunction \eqref{EqS1Solution} becomes
\begin{align}
    \begin{split}
        \Psi \left( g_{tt}, z, \phi_t; k_0, c_0, \epsilon_0, \mu_0 \right) = & \delta \left( g_{tt} = - \frac{f(z)}{z^2} e^{-2k_0} \right) \delta \left( \phi_t - \mu_0 = c_0e^{-k_0} z \right) \\
        & \times \exp \left \lbrace - i \left( \epsilon_0 - 4 e^{-k_0} /L^2 z^3  \right) + i c_0 \mu_0 \right \rbrace,  
    \end{split}
\label{EqClassicalWavefunction}
\end{align}
where
\begin{align}
    f(z) = \frac{1}{L^2} - \frac{z^3}{2} \epsilon_0 e^{k_0} + \frac{c_0^2}{4}z^4.
\end{align}
Here, as in \cite{Blacker:2023oan}, we have taken the limit in which the strongly peaked Gaussian that arises after integration becomes a delta function, localising our state on the classical solution. We have used $z=1/\sqrt{g_{xx}}$ to make explicit the connection with \eqref{EqClassicalsol}. The phase in \eqref{EqClassicalWavefunction} is equal to the onshell action, up to an additional $c_0\mu_0$ term which could be incorporated into our choice of $\beta(c)$ without affecting the classical dynamics.\\\\
One could complete a similar computation in the basis $\lbrace \epsilon, \mu \rbrace$, by evaluating \eqref{EqFourierSolution} on the Fourier transform of \eqref{EqGaussianWavepacket}
\begin{align}
    \alpha \left( \epsilon, f \right) = N_{\alpha} \exp \left \lbrace i k_0 \epsilon - \frac{\left( \epsilon - \epsilon_0 \right)^2}{2 \Delta_k^2} \right \rbrace \exp \left \lbrace i c_0 \mu - \frac{\left( \mu - \mu_0 \right)^2}{2 \Delta_c^2} \right \rbrace,
\end{align}
where $N_{\alpha} = 1/\left( 4\pi \Delta_c \Delta_k \right)$. To obtain the classical solution \eqref{EqClassicalWavefunction}, we require these $\alpha \left( \epsilon,\mu \right)$ wavepackets to be strongly peaked on $\lbrace \epsilon = \epsilon_0, \mu = \mu_0 \rbrace$, and thus that $\epsilon_0 \gg \Delta_k$ and $\mu_0 \gg \Delta_c$. That is, the semiclassical regime is obtained by requiring 
\begin{align}
    \epsilon_0 \gg \Delta_k \gg 1, \text{ and } \mu_0 \gg \Delta_c \gg 1.
\label{EqSemiclassicalConstraint}
\end{align}
Physically, this is because if the wavepacket is to strongly localised on any coordinate (i.e. $k$ or $c$) the variance in its conjugate momentum (i.e. $\epsilon$ or $\mu$) will become large. The constraint \eqref{EqSemiclassicalConstraint} is therefore necessary to preserve the classicallity of the minisuperspace, as will be discussed in Section \ref{SectionClocks}.
\section{Clocks and Expectation Values}
\label{SectionClocks}
The Wheeler DeWitt equation is timeless in the sense that it contains no explicit time parameter, but provides a relation between metric functions on any slicing of spacetime. To specify our slicing is to treat some coordinate (or some combination of coordinates) as constant on each slice. Moving between slices is equivalent to evolving that coordinate. That is, that coordinate acts as a clock, and the Wheeler DeWitt equation can then be used to compute probabilities which are conditional upon the choice of clock. These probabilities can be used to compute expectation values to validate the stability of our semiclassical theory.\\\\
A number of different possible clocks where considered for the AdS-Schwarzschild black hole in \cite{Hartnoll:2022snh}. The $g_{tt}$ clock turned out to be convenient because a Hamilton-Jacobi function linear in $g_{tt}$ was obtained. In that work, evolving $g_{tt}$ from $-\infty$ to $\infty$ was equivalent to evolving the wavefunction from the AdS boundary ($g_{tt} = - \infty$), through the horizon ($g_{tt} = 0$), and to the singularity ($g_{tt} = \infty$). However, as discussed in Section \ref{SecRNAdSHJ}, when the black hole is imbued with a charge, $g_{tt}$ is no longer monotonic from the boundary to the singularity. Therefore, although \eqref{solution} is also linear in $g_{tt}$, $g_{tt}$ may not be the most convenient choice of clock. The fact that $g_{tt}$ is not monotonic corresponds to the presence of two horizons, which also occurs in the dS-Schwarzschild black hole studied in \cite{Blacker:2023oan}. In that work, constant $R$ slices were used to move from the cosmological to the black hole horizons. \\\\
The analogous choice here, as advertised in Section \ref{SecRNAdSHJ}, is to choose $g_{xx}$ as our clock. This is because $g_{xx}$ is monotonic from the AdS Boundary ($g_{xx} \rightarrow \infty$) to the singularity ($g_{xx} \rightarrow 0$). It is straightforward to show that $g_{xx}$ is a null direction in the minisuperspace we are considering here. This means to take $g_{xx}$ as a clock is to define the conserved norm via a limiting sequence of spacelike slices. That is, we can compute conditional probabilities for any given $g_{xx}$, by defining a conserved norm as in \cite{PhysRev.160.1113}. For $g_{xx}$, this norm is
\begin{align}
    \left| \Psi \right|_{g_{xx}}^2 = - \frac{i}{2} \int d\phi_t \int dg_{tt} \left( \Psi^* \partial_{g_{tt}} \Psi - \Psi \partial_{g_{tt}} \Psi^* \right). 
\label{EqGXXNorm}
\end{align}
One can show using the Wheeler DeWitt equation \eqref{EqWdWEquation} that $\partial_{g_{xx}} \left| \Psi \right|_{g_{xx}}^2 = 0$, provided $\Psi$ decays at large and small $\lbrace \phi_t, g_{tt} \rbrace$ on each $g_{xx}$ slice. That is, the norm is conserved under evolution between different slices of fixed $g_{xx}$. Evaluating this norm on a general semiclassical state of the form \eqref{EqS1Solution}, we find
\begin{align}
    \left| \Psi \right|_{g_{xx}}^2 = \int \frac{dkdc}{(2\pi)^2} \left| \beta(k,c) \right|^2.
\end{align}
One could alternatively compute the norm on the Fourier transformed state \eqref{EqFourierSolution}, however the $\lbrace k,c \rbrace$ basis is a more computationally convenient choice.\\\\
An upshot of having defined a norm is that we can now compute expectation values, which can be used to examine the stability of minisuperspace classicality in the semi-classical regime. The simplest, non-trivial expectation values are those of the unfixed metric functions $\langle \phi_t \rangle$ and $\langle g_{tt} \rangle$, which we evaluate on \eqref{EqS1Solution} to obtain respectively
\begin{align}
    \begin{split}
        \langle \phi_t \rangle = & - \frac{i}{2} \int d\phi_t \phi_t \int dg_{tt} \left( \Psi^* \partial_{g_{tt}} \Psi - \Psi \partial_{g_{tt}} \Psi^* \right) \\
    = & \frac{i}{2} \int \frac{dk dc}{(2\pi)^2} \left[ \beta^* \partial_c \beta - \beta \partial_c \beta^* \right] + \int \frac{dkdc}{(2\pi)^2} \frac{e^{-k}c}{\sqrt{g_{xx}}} \left| \beta \right|^2,
    \end{split}
\label{EqGAexpection}
\end{align}
and
\begin{align}
    \begin{split}
        \langle g_{tt} \rangle = & - \frac{i}{2} \int d\phi_t \int dg_{tt} g_{tt} \left( \Psi^* \partial_{g_{tt}} \Psi - \Psi \partial_{g_{tt}} \Psi^* \right) \\
        = & -\int \frac{dkdc}{(2\pi)^2} \left( \frac{i}{2} \left[ \beta \partial_k \beta^* - \beta^* \partial_k \beta \right] \frac{e^{-k}}{2\sqrt{g_{xx}}} + \left| \beta \right|^2 \frac{c^2+4g_{xx}^2/L^2}{4g_{xx}} e^{-2k} \right).
    \end{split}
\label{EqGTTexpectation}
\end{align}
Note here we have left the $\lbrace k,c \rbrace$ dependence implicit. If we evaluate \eqref{EqGAexpection} on the Gaussian wavepacket \eqref{EqGaussianWavepacket}, we obtain 
\begin{align}
    \langle \phi_t \rangle - \mu_0 = \frac{c_0 e^{-k_0}}{\sqrt{g_{xx}}} \exp \left( \frac{1}{4\Delta_k^2} \right) = \frac{c_0 e^{-k_0}}{\sqrt{g_{xx}}} + O \left( 1 / \Delta_k^2 \right),
\label{EqGAsemiclassical}
\end{align}
where expanding in $\Delta_k \gg 1$ we obtain the classical solution \eqref{EqClassicalsol} as the leading contribution. Here, we have included the full quantum correction, to make explicit that the $\mu_0$ contribution is independent of the quantum fluctuations. Thus, the freedom to redefine $\phi_t \rightarrow \phi_t - \mu_0$ is preserved. We can similarly evaluate \eqref{EqGTTexpectation} on the Gaussian wavepacket \eqref{EqGaussianWavepacket}, expanding in $\Delta_k, \Delta_c \gg 1$ to obtain
\begin{align}
    \langle g_{tt} \rangle = -\frac{e^{-2k_0}}{g_{xx}} \left( g_{xx}^2 - \frac{1}{2} e^{k_0} \sqrt{g_{xx}} \epsilon_0  + \frac{c_0^2}{4} \right) + O \left( \frac{1}{\Delta_c^2}, \frac{1}{\Delta_k^2} \right), 
\end{align}
again obtaining to leading order the classical solution \eqref{EqClassicalsol}.\\\\
We can also compute the expectation values of momenta. For example, if $g_{xx}$ is our clock defining our notion of time, $\pi_{xx} = - i \partial_{g_{xx}}$ generates `time translations' and therefore defines the `Hamiltonian' for this clock. We compute the expectation value on the state \eqref{EqS1Solution} to obtain
\begin{align}
    \begin{split}
        \langle \pi_{xx} \rangle = & -\frac{1}{2} \int d\phi_t \int dg_{tt} \left( \Psi^* \partial_{g_{xx}} \left( \partial_{g_{tt}} \Psi \right) - \partial_{g_{xx}} \Psi \partial_{g_{tt}} \Psi^* \right) \\
        = & \int d\phi_t \int dg_{tt} \left[ \frac{g_{tt}}{2 g_{xx}} \left| \frac{\partial \Psi}{\partial g_{tt}} \right|^2 + \frac{1}{2g_{xx}} \left| \frac{\partial \Psi}{\partial \phi_t} \right|^2 - \frac{6}{L^2} g_{xx} \left| \Psi \right|^2 \right] \\
        = & \int \frac{dk dc}{(2\pi)^2} \left( \frac{i}{4g_{xx}} \left( \beta^* \partial_k \beta - \beta^* \partial_k \beta' \right) - 4 e^{-k} \sqrt{g_{xx}}/L^2 \left| \beta \right|^2 \right).
    \end{split}
\label{EqPiXXGeneral}
\end{align}
To obtain the second line we used the Wheeler DeWitt equation \eqref{EqWdWEquation}. Evaluating \eqref{EqPiXXGeneral} on the Guassian wavepacket we obtain
\begin{align}
    \langle \pi_{xx} \rangle = \frac{\epsilon_0}{2g_{xx}} - 4 e^{-k_0} \sqrt{g_{xx}}/L^2 \exp \left( \frac{1}{\Delta_k^2} \right) = \frac{\epsilon_0}{2g_{xx}} - 4 e^{-k_0} \sqrt{g_{xx}}/L^2 + O \left( 1/ \Delta_k^2 \right),
\end{align}
where expanding in $\Delta_k \gg 1$ we obtain the classical solution as the leading contribution, as from \eqref{solution} $\pi_{xx} = \partial S_1/\partial g_{xx} = \epsilon_0/(2g_{xx}) - 4 e^{-k_0} \sqrt{g_{xx}}/L^2$. For $\epsilon_0 >0$, $\langle \pi_{xx} \rangle$ is monotonic. That the expansion does not break down anywhere indicates we have picked a good clock for extending our Wheeler DeWitt solutions through the spacetime.\\\\
Recall that the motivation for introducing these expectation values was to study the stability of minisuperspace classicality as the singularity is approached as $g_{xx} \rightarrow 0$. A natural quantity to test that is the fluctuations of the metric function $g_{tt}$. Computing the variance of $g_{tt}$ we obtain
\begin{align}
    \langle g_{tt}^2 \rangle - \langle g_{tt} \rangle^2 = \frac{\Delta_k^2}{e^{2k_0}8g_{xx}} + O \left( \frac{1}{\Delta_k^2}, \frac{1}{\Delta_c^2} \right),
\label{EqgttVariance}
\end{align}
where we have used $\Delta_k, \Delta_c \gg 1$. One may recognise the denominator as (to leading order) $\langle \pi_{tt} \rangle^2 = 4e^{2k_0}g_{xx}$, and indeed it is straightforward recover the uncertainty relation
\begin{align}
    \left( \langle \pi_{tt}^2 \rangle - \langle \pi_{tt} \rangle^2 \right) \left( \langle g_{tt}^2 \rangle - \langle g_{tt} \rangle^2 \right) = \frac{1}{4}  + O \left( \frac{1}{\Delta_k^2}, \frac{1}{\Delta_c^2} \right).
\end{align}
To interpret \eqref{EqgttVariance} near the singularity it will be useful to remove the $g_{xx}$ dependence. The leading order dependence of $g_{tt}$ on $g_{xx}$ depends on whether or not charge is present. In the neutral ($c_0 = 0$) case
\begin{align}
    \left. \lim_{g_{xx}\rightarrow 0} \langle g_{tt} \rangle \right|_{c_0 = 0} =  \frac{\epsilon_0 e^{-k_0}}{2\sqrt{g_{xx}}} + O \left( g_{xx}, \frac{1}{\Delta_k^2}, \frac{1}{\Delta_c^2} \right),
\label{Eqgttlimitneutral}
\end{align}
so the singularity is spacelike as $\langle g_{tt} \rangle > 0$. Using \eqref{Eqgttlimitneutral} we find that in the neutral case
\begin{align}
    \left. \lim_{g_{xx}\rightarrow 0} \langle g_{tt}^2 \rangle - \langle g_{tt} \rangle^2 \right|_{c_0 = 0} = \frac{\Delta_k^2}{2 \epsilon_0^2} \langle g_{tt} \rangle^2 + + O \left( \frac{1}{\Delta_k^2}, \frac{1}{\Delta_c^2} \right).
\label{EqVarianceNoCharge}
\end{align}
Recall that in the semiclassical regime \eqref{EqSemiclassicalConstraint} $\epsilon_0 \gg \Delta_k$, so the variance is small compared to the expectation value and the Gaussian wavepacket is able to probe the singularity, as noted in \cite{Hartnoll:2022snh}.\\\\
When the black hole has a charge (i.e. $c_0 \not = 0$), the leading order dependence of $g_{tt}$ on $g_{xx}$ becomes
\begin{align}
    \lim_{g_{xx}\rightarrow 0} \langle g_{tt} \rangle = - \frac{c_0^2 e^{-2k_0}}{4g_{xx}} + O \left( \frac{1}{\sqrt{g_{xx}}}, \frac{1}{\Delta_k^2}, \frac{1}{\Delta_c^2} \right).
\label{Eqgttlimitcharge}
\end{align}
As is to be expected from the classical solution \eqref{EqClassicalsolinz}, we see that in the charged case the singularity is timelike as $\langle g_{tt} \rangle < 0$. Note that by taking the $g_{xx} \rightarrow \infty$ limit, we have continued our wavepacket through the inner Cauchy horizon. This inner horizon is believed to be unstable \cite{balasubramanian_holographic_2020,papadodimas_simple_2020,kehle_blowup_2021}, however this instability does not arise in any of the quantities we have computed in our minisuperspace description. Hence, it appears safe to continue the wavepacket through the Cauchy horizon.\\\\
Using \eqref{Eqgttlimitcharge}, we compute
\begin{align}
    \lim_{g_{xx}\rightarrow 0} \langle g_{tt}^2 \rangle - \langle g_{tt} \rangle^2 = -\frac{\Delta_k^2}{2 c_0^2} \langle g_{tt} \rangle + O \left( \frac{1}{\Delta_k^2}, \frac{1}{\Delta_c^2} \right).
\label{Eqgxxvariancecharge}
\end{align}
In \eqref{Eqgxxvariancecharge} the sign of the variance is still positive, as from \eqref{Eqgttlimitcharge} we have that $\langle g_{tt} \rangle < 0$ in the limit $g_{xx} \rightarrow 0$. The significance of \eqref{Eqgxxvariancecharge} is that the variance in $g_{tt}$ is linear in $g_{tt}$ rather than quadratic when $\Delta_k, \Delta_c \gg 1$. If we divide by $\langle g_{tt} \rangle^2$, we see that for any choice of $\Delta_k$ and $c_0$ the variance will go to zero at the singularity. That means the quantum fluctuations are not enough to prevent this direction from collapsing as $g_{xx} \rightarrow 0$, and the classicality of the minisuperspace approximation breaks down. Therefore, the Wheeler DeWitt state is not able to resolve the singularity.\\\\
One possible interpretation is that this result is because the singularity is timelike. With that in mind, it is useful to recall the strong cosmic censorship conjecture: that from physically reasonable generic initial conditions, only spacelike or null singularities can form classically \cite{penrose_notitle_1979}. There has been some evidence \cite{balasubramanian_holographic_2020} via holography that quantum effects enforce strong cosmic censorship in charged AdS black holes. The result \eqref{Eqgxxvariancecharge} may therefore indicate the quantum corrections accounted for by our semiclassical picture are similarly enforcing strong cosmic censorship. It would be interesting to perform computations analogous to \eqref{Eqgxxvariancecharge} for Wheeler DeWitt states in other spacetimes which contain a timelike singularity, to see if a similar result is obtained.\\\\
Throughout these calculations, we have considered $g_{xx}$ as a monotonic function from which to define a clock. As noted in Section \ref{SecRNAdSHJ}, $\phi_t \propto 1/\sqrt{g_{xx}}$ is also monotonic from the AdS boundary ($\phi_t \rightarrow 0$) to the singularity ($\phi_t \rightarrow \infty$). Indeed, one may have instead chosen to define $\phi_t$ as a clock, with an associated conserved norm
\begin{align}
    \left| \Psi \right|_{\phi_t}^2 = \frac{i}{2} \int dg_{xx} \int dg_{tt} \frac{1}{g_{xx}} \left( \Psi^* \partial_{\phi_t} \Psi - \Psi \partial_{\phi_t} \Psi^* \right).
\end{align}
In that case, one can show that on states of the form \eqref{EqS1Solution}
\begin{align}
    \left| \Psi \right|_{\phi_t}^2 = \int \frac{dk dc}{(2\pi)^2} \left| \beta (k,c) \right|^2.
\end{align}
That is, the $g_{xx}$ and $\phi_t$ clocks lead to the same norm. This is to be expected, given the inverse proportionality of the functions. However, the integrals used to compute expectation values in the $\phi_t$ norm are not as easy to evaluate as when using the $g_{xx}$ clock, hence our preference for that choice in this work.
\section{The Boundary Partition Function}
\label{SecPartitionFunction}
From the AdS/CFT correspondence, we expect our theory of the RN-AdS bulk to be dual to some quantum theory living on the AdS boundary. In particular, the semiclassical bulk wavefunctions \eqref{EqS1Solution} constructed from \eqref{solution} should be related to some partition function living on the AdS boundary $g_{xx} \rightarrow \infty$. As noted in the introduction, the Hamilton-Jacobi function $S$ should control the holographic renormalization group flow, with the arguments of $S$ playing the role of coupling in the dual quantum theory. Here, we make that link explicit by studying the energy flow of the dual theory, and show that by running the coupling $g_{xx}$ we move our partition function into the bulk and obtain a Wheeler DeWitt state. Equivalently, by evolving the Wheeler DeWitt state along $g_{xx}$ slices to the boundary, we recover the partition function. \\\\
The holographic framework we consider is analogous to \cite{Hartnoll:2022snh}. Specifically, the standard holographic relation tells us that the partition function of the Lorentzian quantum field theory (QFT) on the AdS boundary is
\begin{align}
    Z_{\text{QFT}} \left[ \gamma, A_{\gamma} \right] = \int \mathcal{D}g \mathcal{D}A e^{iI[g,A]+i S_{\text{ct}}[\gamma]},
\label{EqDictionary}
\end{align}
where the path integrals are over bulk metrics $g$ that are asymptotically AdS with conformal boundary metric $\gamma$, and over vector potentials $A$. $A_{\gamma}$ is the value of the Maxwell field at the boundary. In our minisuperspace approximation, $\gamma_{tt}= g_{tt}$ and $\gamma_{xx} = \gamma_{yy} = g_{xx}$. Here, $I[g,A]$ is the bulk action and $S_{\text{ct}} [\gamma]$ a boundary counterterm action to cancel the divergence in the action at the boundary \cite{Balasubramanian:1999re,Emparan:1999pm}
\begin{align}
    S_{\text{ct}}[\gamma] = 4 \int dx^3 \frac{\sqrt{-\gamma}}{\Delta t \left( \Delta x \right)^2 L} = \frac{4 \sqrt{-g_{tt}}g_{xx}}{L}.
\label{Eqcounterterms}
\end{align}
A difference between our setup and that of \cite{Hartnoll:2022snh} is, as motivated by Section \ref{SectionClocks}, we will use $g_{xx}$ as our clock. The AdS boundary lives at $g_{xx} \rightarrow \infty$, and we will consider the holographic renormalization group flow that evolves as a function of $g_{xx}$. That is, we start with a QFT for a finite $g_{xx}$, and then obtain a conformal field theory (CFT) in the limit that $g_{xx} \rightarrow \infty$. As in \cite{Hartnoll:2022snh}, there is no need to remove a conformal factor from the metric, as conformal invariance at the AdS boundary constrains observables here. \\\\
If we wish to make more explicit our definition of the boundary partition function in terms of a trace over states in the boundary theory, we should first understand how imbuing the black hole with charge affects the dual QFT. The leading order contribution to \eqref{EqDictionary} is the classical solution, upon which $I[g,A]$ is precisely the Hamilton-Jacobi function. In Section \ref{SecRNAdSHJ}, we found several classically equivalent Hamilton-Jacobi solutions, which are each described by a different choice of constants from each pair $\lbrace k_0, \epsilon \rbrace$ and $\lbrace c_0 , \mu \rbrace$. To determine which of these will be most convenient for describing our dual theory, we will take advantage of the AdS/CFT correspondence and draw some intuition from the bulk. Classically, the RN-AdS bulk is described by its mass and charge, which would suggest that $\epsilon$ and $c_0$ should specify the dual quantum theory. We therefore want their conjugate variables $k_0$ and $\mu$ to be variables in our dual partition function, so \eqref{EqSkf} is a useful form of the Hamilton-Jacobi function for us to use. Thus, with the addition of the counterterms \eqref{Eqcounterterms}, we obtain 
\begin{align}
    \begin{split}
        \log Z_{\text{QFT}} \left[ g_{tt}, g_{xx}, \phi_t; k_0,\mu_0 \right] = & - 2i e^{-k_0} \sqrt{g_{xx}}\left( e^{k_0} \sqrt{-g_{tt}}- \sqrt{g_{xx}}/L \right)^2 \\
        & - \frac{i}{2} e^{k_0} \left( \mu_0^2 - \phi_t^2 \right) \sqrt{g_{xx}}.
\label{EqPartionfunctionf0}
    \end{split}
\end{align}
We see explicitly that this partition funciton depends not only on the gauge field $\phi_t$ and the boundary data $g_{tt}$ and $g_{xx}$, but additionally the classical parameters $k_0$ and $\mu_0$. It obeys the scale invariance
\begin{align}
    \log Z_{\text{QFT}} [g_{tt}e^{-\frac{4}{3}k},g_{xx}e^{\frac{2}{3}k},\phi_te^{-\frac{2}{3}k};\mu_0 e^{-\frac{2}{3}k},k_0+k] = \log Z_{\text{QFT}} [g_{tt},g_{xx},\phi_t;\mu_0,k_0].
\label{EqZInvariancef0}
\end{align}
The energy density of the dual field theory is defined in the usual way, by the momentum conjugate to $g_{tt}$
\begin{align}
    \sqrt{-\gamma}\langle T^t_t\rangle_{\text{QFT}}=-2i\gamma_{tt}\frac{\partial \log Z_{\text{QFT}}}{\partial \gamma_{tt}}=4\sqrt{-g_{tt}g_{xx}}(\sqrt{g_{xx}}/L-e^{k_0}\sqrt{-g_{tt}}).
\end{align}
To obtain the CFT limit, we eliminate $g_{tt}$ using \eqref{EqClassicalsol}, and expand in $g_{xx} \rightarrow \infty$ to obtain
\begin{align}
    \lim_{g_{xx} \rightarrow \infty}\sqrt{-\gamma}\langle T^t_t\rangle_{\text{QFT}} = \epsilon.
\label{EqEnergyDensityBoundary}
\end{align}
We recover, as in \cite{Hartnoll:2022snh}, that the energy density of the black hole $\epsilon$ is the energy density of the dual theory. Indeed, from \cite{Hartnoll:2022snh} we expect the conjugate variable $k_0$ to play a role anaologous to time in the CFT limit. To learn more about our dual theory, it will be instructive to study the dependence of $\log Z_{\text{QFT}}$ on $k_0$. In particular, observe that
\begin{align}
    \begin{split}
        - i \frac{\partial \log Z_{\text{QFT}}}{\partial k_0} = &  \frac{1}{2} \sqrt{g_{xx}} \left(e^{k_{0}} \left(-f_{0}^2+\phi_t^2+4 g_{tt}\right)+4 g_{xx} e^{-k_0}/L^2 \right) \\
        = & \epsilon + c_0 \mu_0 \\
        = & \epsilon - \mu_B Q.
    \end{split}
\label{EqDLogDk0boudnary}
\end{align}
In the second line we have used the classical solutions for $g_{tt}$ and $\phi_t$. We recognise from \eqref{EqEnergyDensityBoundary} that $\epsilon$ is the energy density in the $g_{xx} \rightarrow \infty$. In the third line we replaced $c_0 \mu_0$ with the boundary charge and chemical potential from \eqref{EqMuandRho} and \eqref{EqChargeDefinition}. From \eqref{EqDLogDk0boudnary} it appears that our dual theory is characterised by an energy and a charge. That is, beyond the classical regime we expect
\begin{align}
    - i \frac{\partial \log Z_{\text{QFT}}}{\partial k_0} = H_{\text{QFT}} - \mu_B Q_{\text{QFT}}.
\label{EqDlokDk0semi}
\end{align}
Here, $H_{QFT} = \epsilon$ and $Q_{\text{QFT}} = Q$ in the classical limit. We therefore expect
\begin{align}
    Z_{\text{QFT}} [g_{tt},g_{xx},\phi_t;k_0,f_0] = \text{Tr} \left( e^{i k_0 \left( H_{QFT}[g_{tt},g_{xx},\phi_t]- \mu_B Q_{QFT} \right)} \right),
\label{EqClaim}
\end{align}
to be a suitable partition function of the boundary theory. \\\\
All of our discussion so far has only considered the classical limit of \eqref{EqDictionary}. To further justify our claim \eqref{EqClaim}, we should consider \eqref{EqDictionary} beyond the classical regime. Recall that the path integral over $e^{iI[g,A]}$ is usually associated with the wavefunction of the gravititational bulk \cite{PhysRevD.28.2960}. Therefore, when the boundary counterterms are subtracted out, in the semiclassical regime the partition function \eqref{EqDictionary} should obey the Wheeler DeWitt equation \eqref{EqWdWEquation}. In particular, this means that in the semiclassical regime we can approximate the partition function as 
\begin{align}
    Z_{\text{QFT}} [g_{tt},g_{xx},\phi_t; \beta] = e^{4i\sqrt{-g_{tt}}g_{xx}/L} \Psi \left(g_{tt},g_{xx},\phi_t; \beta \right),
\label{EqPartitionfunctionSemiclass1}
\end{align}
where $\Psi$ is a solution to \eqref{EqWdWEquation}. Note that if we rearrange \eqref{EqPartitionfunctionSemiclass1} for $\Psi$, that is
\begin{align}
    \Psi \left(g_{tt},g_{xx},\phi_t; \beta \right) = e^{-4i\sqrt{-g_{tt}}g_{xx}/L} Z_{\text{QFT}} [g_{tt},g_{xx},\phi_t; \beta],
\label{EqZdefromed}
\end{align}
we could reinterpret the Wheeler DeWitt solution as a deformation of a QFT partition function. Or indeed, treating $g_{xx}$ as a clock, as a deformation of a QFT living on the AdS boundary as we evolve away from the $g_{xx} \rightarrow \infty$ limit. This interpretation could be linked to studies of $T^2$ deformations of the boundary theory in AdS/CFT \cite{Belin:2020oib,LeFloch:2019rut,McGough:2016lol}, or indeed the explicit specification of a bulk state in terms of a boundary theory in \cite{Araujo-Regado:2022gvw}.\\\\
To explicitly compute \eqref{EqPartitionfunctionSemiclass1}, we need to use a Wheeler DeWitt state $\Psi$. In Section \ref{SecConstructingStates}, we discussed how to construct a Wheeler DeWitt state \eqref{EqS1Solution} from a basis of solutions \eqref{EqBasisS1} weighted by some function $\beta(k,c)$. What we want to do here is slightly different; as discussed earlier, \eqref{EqSkf} is a more natural representation Hamilton-Jacobi function to identify with our partition function. We will therefore perform a Fourier transform to the $c$ integral in \eqref{EqS1Solution} to recast our states in the semiclassical basis $e^{iS_{k,\mu}}$. To do so, we will assume that our wavepackets are separable. That is, $\beta(k,c) = \beta_k(k) \beta_c(c)$ and $\alpha (\epsilon,\mu) = \alpha_{\epsilon} (\epsilon) \alpha_{\mu} (\mu)$. Performing the fourier transform by stationary phase, we recover
\begin{align}
    \Psi \left( g_{tt}, g_{xx}, \phi_t \right) \sim \int dk d\mu \beta_k(k) \alpha_{\mu}(\mu) e^{iS_{g_{tt},g_{xx},\phi_t;k,\mu}}.
\label{EqinKF}
\end{align}
Here, $\sim$ means up to some prefactor. Therefore, in the semiclassical regime, from \eqref{EqPartitionfunctionSemiclass1} and \eqref{EqinKF} our claim \eqref{EqClaim} is that the partition function becomes
\begin{align}
    Z_{\text{QFT}} \left[ g_{tt}, g_{xx}, \phi_t; \beta \right] \sim \int dk df \beta_k(k) \alpha_{\mu}(\mu) \text{Tr} \left( e^{i k \left( H_{\text{QFT}}- \mu \rho_{\text{QFT}} \right)} \right).
\label{EqClaim2}
\end{align}
Note that here we use the charge density $\rho_{\text{QFT}} = e^{k_0} Q_{\text{QFT}}$. This is because it is more natural to build our wavepackets in $f$ than $\mu$, given the arguments of the Hamilton-Jacobi function. For the Gaussian wavepacket \eqref{EqGaussianWavepacket}, \eqref{EqClaim2} becomes
\begin{align}
    Z_{\text{QFT}} \sim \text{Tr}_{\epsilon,c}\left(e^{-(H_{\text{QFT}}-\epsilon_0)^2/(2\Delta_k^2)-\Delta_c^2(c-c_0)^2/2}e^{i k_0 \left( H_{QFT}- \mu_0 \rho_{QFT} \right)}\right). 
\label{EqClaim3}
\end{align}
We can now assess the validity of the claim \eqref{EqClaim} by explicitly evaluating \eqref{EqPartitionfunctionSemiclass1} and comparing it to \eqref{EqClaim3}. We evaluate \eqref{EqPartitionfunctionSemiclass1} by making a stationary phase approximation and taking the $g_{xx} \rightarrow \infty$ limit to obtain
\begin{align}
    \begin{split}
        \lim_{g_{xx} \rightarrow \infty} Z_{\text{QFT}} \sim \int d\epsilon dc & \exp \left \lbrace - \frac{\Delta_c^2}{2} \left( c- c_0 \right)^2 - \frac{\left( \epsilon - \epsilon_0\right)^2}{2\Delta_k^2} + i \epsilon \log \left( e^{k_0} L \frac{\sqrt{-g_{tt}}}{\sqrt{g_{xx}}} \right) \right \rbrace \\
        & \times \exp \left \lbrace i \left( c \phi_t + \frac{\epsilon^2 L}{8 \sqrt{-g_{tt}}g_{xx}} + \frac{c^2 L \sqrt{-g_{tt}}}{2g_{xx}}  \right)  \right \rbrace,
    \end{split}
\label{EqPartitionfunctionSemiclass3}
\end{align}
up to a prefactor which is not important here. The integrals in \eqref{EqPartitionfunctionSemiclass3} are identified with the trace in \eqref{EqClaim3}, and the Gaussian terms are identical. The third term in the first line of \eqref{EqPartitionfunctionSemiclass3} is due to the smearing of $k$ around $k_0$ and changes the sign of the $\epsilon^2$ term, as noted in the neutral case \cite{Hartnoll:2022snh}. The term linear in $\phi_t$ arises from an analogous blurring of $\mu$ around $\mu_0$, and also changes the sign of the $c^2$ term. As in the neutral case, the $\epsilon^2$ term is the scaling for the density of states in a $2+1$ dimensional CFT. To interpret the $c^2$ term, we compute the current associated with the electromagnetic potential $A$. Classically, in the $dt$ direction this is
\begin{align}
    J_t = \frac{\pi_{\phi}^2}{2g_{xx}^2},
\end{align}
so the $c^2$ term in \eqref{EqPartitionfunctionSemiclass3} is identified with the energy density $\sqrt{-\gamma} \langle J_t \rangle$ of the Maxwell field. Finally, to interpret the term linear in $\phi_t$, we compute 
\begin{align}
    -i \lim_{g_{xx}\rightarrow \infty} \frac{\partial \log Z_{\text{QFT}}}{\partial \phi_t} = \langle c \rangle = - \langle \rho \rangle.
\label{EqSemiZCharge}
\end{align}
Here, we have an expectation value because we are taking a trace of the boundary theory ensemble. The implication of \eqref{EqSemiZCharge} is that the bulk gauge field is dual to the charge density in the bulk theory. This recovers the standard result of the holographic dictionary \cite{hartnoll_holographic_2018}. We have therefore identified each term in \eqref{EqPartitionfunctionSemiclass3} with those in \eqref{EqClaim3}, validating our proposition for the dual theory ensemble.\\\\
To summarise, what we have done here is demonstrated at a classical and semiclassical level that Wheeler DeWitt states of the RN-AdS interior can be constructed from the holographic renormalization group flow of $g_{xx}$ of a partition function that lives on the AdS Boundary. The gaussian wavepacket localises to a fixed energy and charge, suggesting that the partition function of the dual theory is specified by those quantities.
\section{Black hole thermodynamics}
\label{SecThermodynamics}
Having identified the dual theory as being specified by a fixed energy and charge, we expect to be able to construct a similar thermodynamic picture for the bulk theory. In particular, we expect that at the black hole horizon, Wheeler DeWitt states should know about the black hole thermodynamics. That quantum states of a gravitational bulk should know about horizon thermodynamics has long been appreciated \cite{gibbons_action_1977}; what we do here is explicitly recover these thermodynamics from our semiclassical states.\\\\
To recover the bulk thermodynamics, we follow a similar procedure to \cite{Blacker:2023oan}. There, it was observed that it is natural to average over a Wheeler DeWitt state such as \eqref{EqClassicalWavefunction} by integrating over a metric function. This averaging has the practical benefit of removing a delta function. Physically, this averaging corresponds to fixing the gravitational gauge redundancy represented by the Wheeler DeWitt equation. In particular, in \cite{Blacker:2023oan} the redundancy was fixed by fixing the trace of the extrinsic curvature $K$, motivated by the observations of \cite{york_role_1972,witten_note_2021,witten_note_2022}. This is achieved by fourier transforming the wavefunction to a basis of extrinsic curvatures rather than metric functions, and then averaging. The trace $K$ is proportional to $\pi_{v}$, the momentum conjugate to the volume $v = \sqrt{-g_{tt}}g_{xx}$. The fourier transform is therefore taking us from wavefunctions $\Psi \left( v, [h], \phi_t \right)$ to $\tilde{\Psi} \left( \pi_v, [h], \phi_t \right)$, where $[h] = g_{tt}/g_{xx}$ is the conformal class of the induced metric. The transform is implemented by adding a term proportional to $i v \pi_v$ in the exponent of \eqref{EqClassicalWavefunction}. As in \cite{Blacker:2023oan}, we will add the Gibbons-Hawking boundary term
\begin{align}
    2 \int d^3x \sqrt{h} K = g_{tt} \pi_{tt} + g_{xx} \pi_{xx},
\end{align}
and consider the partial integration
\begin{align}
    \tilde{\Psi} \left( g_{xx}, \phi_t \right) = \int dg_{tt} e^{-i \left( g_{tt} \langle \pi_{tt} \rangle + g_{xx} \langle \pi_{xx} \rangle \right)} \Psi \left( g_{tt}, g_{xx}, \phi_t \right).
\label{EqPartialIntegrationStep1}
\end{align}
We integrate over $g_{tt}$ because $g_{xx}$ is monotonic from the AdS boundary to the singularity. We will in particular be interested in evaluating \eqref{EqPartialIntegrationStep1} at the black hole horizon $g_{xx} (z_h)$. It is worth noting that in this case, \eqref{EqPartialIntegrationStep1} is equivalent to averaging over a state in a gravitational theory with
\begin{align}
    I_{\text{GR}}\left[ g, A \right]=\int d^4x\sqrt{-g}(R+6/L^2)+2\int d^3x\sqrt{h} \left( K_I - K_h \right),
\end{align}
where $K_I$ and $K_h$ are the trace of the extrinsic curvature at the infinite boundary and the horizon respectively. This is in the spirit of the original Gibbons-Hawking procedure \cite{gibbons_action_1977}. \\\\
Integrating over $g_{tt}$ only removes one of the delta functions from \eqref{EqClassicalWavefunction}. That is, we still have a remaining distribution in $\phi_t$. We proceed by integrating out $\phi_t$ to obtain an effective field theory on a $g_{xx}$ slice to obtain
\begin{align}
    \tilde{\Psi} \left( g_{xx} \right) = \int d\phi_t \int dg_{tt} e^{-i \left( g_{tt} \langle \pi_{tt} \rangle + g_{xx} \langle \pi_{xx} \rangle \right)} \Psi \left( g_{tt}, g_{xx}, \phi_t \right).
\label{EqPartialIntegrationStep2}
\end{align}
Note that we are interested in $\tilde{\Psi}$ at the black hole horizon $g_{xx} = g_{xx,+}$, where $\phi_t = 0$ as per \eqref{EqFixMuB}. We therefore don't lose any information by performing the integral over $\phi_t$. Evaluating \eqref{EqPartialIntegrationStep2} on the solution \eqref{EqClassicalWavefunction} we obtain
\begin{align}
    \begin{split}
        \tilde{\Psi} \left( g_{xx,+} \right) = & \exp \left \lbrace i \left[ e^{-k_0} S T_h - \left( \epsilon + c_0 \mu_0 \right) \right] \right \rbrace \\
         = & \exp \left \lbrace i \left[ e^{-k_0} S_h T_h - \epsilon + \mu_B Q \right] \right \rbrace.
    \end{split}
\label{EqAveragedPsi}
\end{align}
Here we defined the Bekenstein-Hawking entropy
\begin{align}
    S_h = A_h , \text{ where } A_h = \frac{4\pi}{z_h^2} = 4\pi g_{xx,+},  
\end{align}
and the black hole temperature
\begin{align}
    T_h = \frac{1}{4\pi} \left| \frac{\partial f(z)}{\partial z} \right|_{z=z_+} = \frac{1}{4\pi} e^{2k_0} \left| 2\sqrt{g_{xx}} \frac{\partial g_{tt}}{\partial g_{xx}} \right|_{g_{xx}=g_{xx,+}}.
\end{align}
Note that, as in Section \eqref{SecPartitionFunction}, we identified $c_0 \mu_0 = - \mu_B Q$ as before. To proceed further, firstly observe from the metric \eqref{EqFZ} that $\epsilon = M e^{-k_0}$ where $M$ is the mass of the black hole. Additionally, whilst the charge $Q$ in the bulk and boundary is the same, the chemical potential is redshifted when moving from the boundary to the horizon. The redshift is explicitly computed following \cite{bardeen_four_1973}
\begin{align}
    \mu_B = \sqrt{-n_a n^a} \mu_h = e^{-k_0} \mu_h,
\end{align}
where $n^{\mu}$ is the four-vector introduced above \eqref{EqChargeDefinition}. Our averaged wavefunction \eqref{EqAveragedPsi} can therefore be expressed as
\begin{align}
    \Psi \left( g_{xx,+} \right) = & \exp \left \lbrace i e^{-k_0}  \left[ S_h T_h - M + \mu_h Q \right] \right \rbrace = \exp \left \lbrace - i e^{-k_0} W \right \rbrace,
\label{EqAveragedPsi2}
\end{align}
where $W$ is the usual grand canonical thermodynamic potential. In the spirit of \cite{Blacker:2023oan}, if we set $e^{-k_0} = -i/T_h$, the argument of the exponential is the usual thermodynamic argument for Euclidean quantum gravity \cite{gibbons_action_1977}. Setting $e^{-k_0} = -i/T_h$ is equivalent to performing a Wick rotation of the metric \eqref{EqFZ} to Euclidean signature. In Euclidean siganture, the radial coordinate runs from the AdS boundary to the black hole horizon, where the Euclidean time circle shrinks to zero \cite{hartnoll_holographic_2018}. \\\\
It is worth comparing the result \eqref{EqAveragedPsi2} to that in \cite{Blacker:2023oan}, in which a de Sitter-Schwarzschild black hole was considered. There, the averaged and fourier transformed wavepacket recovered exactly the horizon entropy $\tilde{\Psi} = e^{S_h}$ under an analogous Wick rotation. This is because for a closed universe, such as an asymptotically de Sitter one, $W = -TS$, and so $-WT^{-1}$ is the entropy $S$ exactly. We make this comparison to emphasise that it is the thermodynamic potential $W$ which the Wheeler DeWitt state knows about.
\section{Discussion}
In this work, we have studied Wheeler DeWitt solutions as a probe of the interior of charged black holes. In particular, we have applied the framework of \cite{Hartnoll:2022snh} to the RN-AdS black hole, constructing Wheeler DeWitt states in the black hole interior and extending them to the cauchy interior and the exterior. We found that the close to the singularity, quantum fluctuations become significant and the state is unable to probe the singularity. Moving to the exterior and considering the AdS/CFT correspondence, we then found that our Wheeler DeWitt states were part of the renormalization group flow of a quantum theory living on the AdS boundary. The partition function of this dual theory was specified by an energy and charge. We then returned to the bulk description to recover this thermodynamic picture there. In particular, by applying the averaging procedure of \cite{Blacker:2023oan} and making an appropriate Wick rotation, we recovered the grand canonical thermodynamic potential at the black hole horizon. \\\\
Having developed this framework, it is natural to consider what other rich dynamics of charged black holes it will allow us to explore. For example, it was shown in \cite{Hartnoll:2020rwq} that the inner Cauchy horizon of the RN-AdS black hole vanishes upon coupling a scalar field to the bulk theory. Whilst directly coupling a scalar field to our theory has not yet yielded a tractable Hamilton-Jacobi solution, a simpler starting point might be to treat the scalar field as a perturbation of the Hamilton-Jacobi solution as in \cite{halliwell_origin_1985}. Such a treatment would introduce sources in the dual theory living on the AdS boundary and thus affect the renormalization group flow. \\\\
Coupling to a scalar field could equally be used to introduce inhomogeneities around our minisuperspace approximation, which would be a first step to going beyond the semiclassical regime \cite{pimentel_inflationary_2014,Ning:2023ybc}. Going beyond minisuperspace may be useful to better understand the result \eqref{Eqgxxvariancecharge}. That is, it would be interesting to learn whether petrurbations around minisuperspace make it possible for a Wheeler DeWitt state to resolve the timelike singularity. \\\\
Finally, this work should be extended to answer similar questions in de Sitter space. In \cite{Blacker:2023oan}, the Wheeler DeWitt framework of \cite{Hartnoll:2022snh} was applied to recover the dS-Schwarzschild solution and propose a framework for static patch holography. The dS-Schwarzschild solution is another example of a spacetime with two horizons and where $g_{tt}$ is not monotonic, so cannot be used as a clock to probe the interior. Whether the analogous function to $g_{xx}$ in \cite{Blacker:2023oan} ($R$) can be used as a clock may be useful for exploring interior singularities in de Sitter, which are expected to be highly inhomogeneous \cite{belinskii_oscillatory_1971}. Outside the black hole interior, charged de Sitter black holes have interesting properties which Wheeler deWitt solutions could be used to explore semiclassically \cite{morvan_action_2022,castro_near-extremal_2023}.

\acknowledgments
We are especially grateful to Sean Hartnoll for extensive feedback on an early draft of this paper. It is also a pleasure to thank Joseph Conlon, Fernando Quevedo, Ronak M Soni, Aron Wall and Zhenbin Yang for useful conversations. S.N. would like to thank DAMTP at the University of Cambridge and the University of Southampton for their hospitality while the majority of this work was being done. Finally, S.N. would like to specially thank Pam Kivelson and Steven Kivelson for their kind suggestions and warm encouragement. M.J.B. was supported by a Gates Cambridge Scholarship (\#OPP1144). S.N. was supported by the China Scholarship Council-FaZheng Group at the University of Oxford. 



\bibliographystyle{JHEP}
\bibliography{WdWRnAdS}

\end{document}